\documentclass[12pt,thmsa]{article}
\usepackage{sw20aip}

\input{tcilatex}
\begin{document}

\title{REALISTIC INTERPRETATION OF QUANTUM MECHANICS}
\author{Emilio Santos \\
%EndAName
Departamento de F\'{i}sica. Universidad de Cantabria. Santander. Spain}
\maketitle

\begin{abstract}
It is argued that the usual postulates of quantum mechanics are too strong.
It is conjectured that it is possible to interpret all experiments if we
maintain the formalism of quantum theory without modification, but weaken
the postulates concerning the relation between the formalism and the
experiments. A set of postulates is proposed where realism is insured.
Comments on Bell\'{}s theorem are made in the light of the new postulates.
\end{abstract}

\section{\textbf{Introduction}}

After the discovery of quantum mechanics a warm debate took place about its
interpretation, but since 1927 the Copenhagen interpretation (CI), due to
Bohr, dominated in the scientific community and the debate almost ceased,
although a few critical voices remained like Einstein and Schr\"{o}dinger
(see \cite{Wheeler} for reprints of the relevant papers of that period). But
it is interesting that the CI was not understood in the same way by
different people. In particular Bohr did not attempt to apply quantum
mechanics to the macroscopic domain, whilst von Neumann did it \cite{neumann}%
, and even made a model of measurement on this basis. The debate reappeared
with Bohm's work in 1952 and, with more strength, after Bell's paper in 1964 
\cite{bell} and it lasts until today. In the last few decades the CI has
been progresively abandoned and the so-called many worlds interpretation
(MWI) is taken its place, specially amongst cosmologists on the one hand and
workers in quantum information on the other. I include in MWI the
interpretations in terms of decoherence\cite{Zurek} or consistent histories%
\cite{Hartle}, in my opinion they are just (important) elaborations within
MWI. Still, some people claim that no interpretation is really needed \cite
{peres}. The reason for the variety of interpretations of quantum mechanics
is that many predictions of the theory have a paradoxical character, and
people have attemped to solve these paradoxes by different means, without
complete success till now in my opinion.

In my view the first step towards a satisfactory solution of the paradoxes
is to investigate what is the minimal set of postulates of quantum mechanics
which are really indispensable for the interpretation of observations and
experiments. In some sense this approach is what the CI attempted and it is
close to the ''no-interpretation'' above mentioned \cite{peres}. However, at
a difference with the merely instrumentalist (pragmatic, sometimes named
positivistic) character of the CI, I propose including the requirement that
quantum mechanics is universal and realistic. By universal I mean that
quantum mechanics applies to both the microscopic and the macroscopic domain
(although maybe not to the universe as a whole, see below). This contrasts
with the CI (at least Bohr\'{}s) view that the referent of the theory is
always the union of a microscopic system plus a macroscopic context
(including the measuring apparatus), but the macroscopic systems should be
treated according to classical theories.

Realistic means that we assume that \textit{physics} (or science in general) 
\textit{makes assertions about the world and not only about the results of
observations or experiments}. In particular this implies that we shall give
an \textit{ignorance interpretation} to the probabilities predicted by the
theory, at least the probabilities about properties of macroscopic bodies.

In order to expose my proposal I shall divide this paper in five parts with
the following aims:

1) Pointing out that a part of the postulates of quantum theory are
unnecessarily strong,

2) Analyzing some experiments in order to discover what postulates are
really indispensable,

3) Proposing new postulates leading to a minimal realistic interpretation,

4) Studying the relation with other interpretations: Copenhagen, many worlds
and hidden variables,

5) Showing, with the example of Bell\'{}s theorem, how the conceptual
problems are alleviated.

\section{The standard postulates of quantum mechanics}

Any theory of physics contains a (mathematical) formalism plus postulates
giving the connection with observations or experiments (semantical rules).
In quantum mechanics the formalism is the theory of Hilbert spaces combined
with relativistic (Lorentz) invariance plus some particular postulates (e.g.
masses and coupling constants of elementary particles). (Following a common
practice I shall speak about ``quantum mechanics'' in the rest of this
article, but I really mean ``quantum theory''. Also it is known that quantum
fields cannot be formulated in Hilbert spaces, but require the more general
framwork of C* algebras, but I shall ignore this point of mathematical
rigour).

Most textbooks do not attempt to make precise the connection of the
formalism with the observations or experiments, but we might try to divide
the traditional semantical rules in two classes:\ 

1) \emph{The correspondences operators-observables and\ vectors-states}.
They establish that we must associate a vector of an appropriate Hilbert
space to every state of a physical system, and a self-adjoint operator to
every observable. This statement alone gives very little information about
the connection of the formalism with the experiments because no mention is
made of the values of the physical quantities. It should be necessarily
complemented with postulates about the measurement.

2) \emph{The theory of measurement}, which establish that ``when we measure
the observable associated to the operator A in the state with vector
(wavefunction) $\psi $ we obtain one of the eigenvalues of A, say a$_{j}$,
with probability \TEXTsymbol{\vert}$\left\langle \psi \mid \psi
_{j}\right\rangle $\TEXTsymbol{\vert}$^{2}$ ''.

It is increasingly obvious that \emph{we should not postulate }a theory of
mesurement. The measurement is just an interaction between some physical
system and a macroscopic apparatus and therefore the study of the
measurement should be \emph{derived} from the remaining postulates if
quantum mechanics is to be regarded as a fundamental theory of nature. The
problem of including the measurement within the postulates of a theory is
that makes it subjective and ambiguous. Because, what is really a
measurement?, at what time is it made exactly?, a bad experiment, would not
give results in disagreement with those postulated ?. These, and other
arguments, have been brillantly exposed by John Bell, who proposed even the
removal of the word ``measurement'' from physical theories \cite{bell}.
Nevertheless it is a fact that the relation between operators and
observables in quantum mechanics is always stated with reference to the
possible values which may be obtained in measurements. This fact contrast
with the situation in classical physics, for instance classical statistical
mechanics. In that theory we also associate states of physical systems to
some elements of the theory, namely probability distributions in phase
space. Also we associate observables with other elements, namely functions
of the phase space variables. The semantical rules are complete when we
assume that all variables have values simultaneously and give an ignorance
interpretation to the probabilies. But a similar procedure cannot be used in
quantum mechanics as explained in the following.

According to the quantum-mechanical formalism, combined with the standard
correspondence between measurable values of the dinamical variables and the
eigenvalues of the corresponding operators, the variables cannot have a
value when they are not measured. Indeed this is the essential containt of
the Kochen-Specker theorem (see e.g. \cite{redhead} or \cite{mermin}).
Consequently we cannot make statements about observables outside the context
of a measurement, which contradicts the desire of removing the theory of
measurement. I think that the existence of several, quite different,
interpretations of quantum mechanics derives from that difficulty. CI
(Bohr\'{}s) assumes that the connection formalism-experiments should always
involve macroscopic apparatuses which must be treated according to classical
physics. This implies that macroscopic variables \emph{do possess values}
independently of measurement, and the postulates refer to these \emph{%
objective} values. John von Neumann\'{}s interpretation (included in CI by
some people) is actually different. It \emph{does not }attribute values to
macroscopic variables from the start, but assumes that there is a ``collapse
of the wavefunction'' which \emph{objectifies} the values (maybe by the
action of the mind or consciousness of a human observer). MWI solves the
problem with an appeal to many branches of the ``wavefunction of the
universe'', \emph{all relevant variables having values} in each individual
branch. But for me all these interpretations are unsatisfactory. CI, both
Bohr\'{}s and von Neumann\'{}s, because it establish an ``infamous
boundary'' between the macro and the microscopic domain (or between matter
and mind). MWI because it contains assumptions which cannot be tested
empirically and, in addition, look rather bizarre (a copy of each observer
lives in every branch of the universe\'{}s wavefunction). This is why I am
proposing an alternative having elements of both, CI and MWI, but trying to
remove their difficulties.

At this moment a comment of mathematical character is in order. As is well
known, in the standard approach the \emph{states} of physical systems are
associated to \emph{vectors }of the Hilbert space and the \emph{observables}
to \emph{self-adjoing operators.} Actually a generalization is possible if
we associate observables to normalized positive operator valued \emph{(POV)
measures }\cite{busch}. On the other hand, all vectors which are obtained by
multiplication of a given vector times complex numbers are assumed to
represent the same state. Therefore it is appropriate to speak about\emph{\
rays }of the Hilbert space rather than about vectors. Nevertheless, I shall
retain the more common, although mathematically less correct, use of the
words vector and operator because here I am putting the emphasis in the
conceptual, philosophical, questions rather than in the formal,
mathematical, ones.

Let us look more closely to the correspondence \emph{states}-\emph{vectors}
and \emph{observables}-\emph{selfadjoing operators. }They are usually
assumed (explicitly or implicitly) to be \emph{one to one}, except for the
superselection rules. The one-to-one character of the states-vectors
correspondence is sometimes named \emph{superposition principle }and it is
enuntiated ``if two vectors correspond to states, any vector which is a
linear combination of the former also corresponds to a state, except for the
superselection rules''. Actually if the correspondence rays-states is one to
one, also the correspondence \emph{observables}-\emph{selfadjoing operators }%
is one to one and viceversa. (This result is related to Gleason\'{}s theorem 
\cite{redhead}). In my opinion such postulates are unnecessarily strong and
even absurd, because if we assume that every self-adjoint operator
represents an observable, we are making postulates about ``what may be
measured \emph{in principle}, that is what \emph{could} be achieved in the
laboratory in a more or less distant future''. I think that the
correspondence can be only in one direction, and so is stated in carefull
textbooks. That is, we might assume that for every possible state which may
be found in nature or manufactured in the laboratory there is an associated
vector, and that for every observable which can actually be measured there
is a self-adjoint operator, but not viceversa. This may be represented as
follows

\emph{\ states }$\mapsto $ \emph{vectors, observables }$\mapsto $\emph{%
self-adjoint} \emph{operators.}

However, this correspondence is still too strong in my opinion. We should
just admit

\emph{states }$\mapsto $ \emph{density matrices, } \emph{dynamical variables 
}$\mapsto $ \emph{self-adjoint operators }

That is, states should be associated to density operators and only rarely
the density operators would correspond to vectors in Hilbert space, the
so-called pure states . Also we should speak about dynamical variables
rather than observables, because observability is a \emph{practical}
question which should not be postulated for all dinamical variables. A more
precise statement of the postulates which I propose will be made in section
4, but in order to motivate them I shall mention a few typical experiments.

\section{\protect\smallskip The interpretation of experiments}

In the following I consider some experiments in order to see that rather
weak postulates are indispensable:

\textbf{1.Static properties of atoms, nuclei, molecules and solids.}

Probably the most dramatic qualitative and (modulo some unavoidable
approximations due to the complexity of the calculations) quantitive success
of quantum theory is the interpretation of the physics of atoms, nuclei,
molecules and solids, in particular their static properties. For instance,
the prediction of the form, size and binding energy of molecules or crystals
(hence their stability), the electric and magnetic properties (if external
fields are included), etc., is the basis for most of theoretical chemistry
and solid state physics.

In order to get these predictions from quantum mechanics, it is enough to
take into account the evolution of the Schr\"{o}dinger (or
Schr\"{o}dinger-Pauli) equation for electrons and nuclei, coupled to
quantized Maxwell equations for the electromagnetic field. If we impose
appropriate boundary conditions (e.g. no radiation coming from infinity), we
may \emph{derive} the existence of an unique stationary state (``the ground
state''). Therefore, there is no need to\ \emph{postulate} that the ground
state is the eigenstate of the Schr\"{o}dinger equation with the lowest
eigenvalue, this fact following from the formalism. After that, the solution
of the stationary Schr\"{o}dinger equation gives all the required
information, provided we assume that the expectation value of the energy of
the system is given by the standard rule 
\[
E=\left\langle \Psi \left| H\right| \Psi \right\rangle , 
\]
where H is the Hamiltonian operator and $\Psi $ the vector state. In
particular it is not necessary to assume that all self-adjoint operators
represent observables, or that all eigenvectors of the Hamiltonian represent
physical states. Similar arguments apply to the static properties of atoms
or nuclei.

\textbf{2. Collisions}.

For the study of (elastic or inelastic) scattering it is enough to consider
the evolution of two or more systems (atoms, molecules, nuclei,...), both in
the ground state, which initially are at a macroscopic distance and approach
each other. Aside from the quantum evolution equations plus postulates of
classical physics for the interpretation of the final results, we only need
rules for preparation of the initial state and the interpretation of the
final state.

The initial state usually consists of a beam whose preparation may be
described in terms of classical physics (e. g. a macroscopic accelerator).
Hence we should derive the density matrix corresponding to the (usually
microscopic) quantum systems in the incoming beam and the detector.
Nevertheless I do not think that it is possible to propose any general
postulate saying how to do that. We should use the method or trial and error
with the only general rule that that the appropriate density matrix is the
one havint the greatest von Neumann entropy compatible with our information.
After that we must use the quantum formalism in order to compute the
evolution of the initial density matrix until the detector. Then we should
apply\ to the detection process the quantum formalism, which most times
could be approximated by classical equations. At the end we arrive at a
density matrix for the final state of the measuring apparatus. Decoherence
theory\cite{Zurek} shows that the apparatus density matrix is, to a very
good approximation, diagonal in the coordinates representation. That density
matrix is interpreted as a probability distribution (with the ignorance
interpretation.) I think that this is the way how physicists in labs
actually interpret the collision experiments.

This interpretation is very good \textit{for all practical purposes (}FAPP),
but presents a fundamental difficulty (Bell pointed out the important
difference between FAPP and fundamental assertions\cite{bell}). In fact the
final density matrix of the measuring apparatus is only approximately, but
not exactly, diagonal. Therefore either we renounce to the interpretation of
its diagonal elements as true probabilities (this should be the position of
MWI) or we break quantum mechanics at the macroscopic level (this would be
the position of CI). My position departs from both MWI and CI by assuming
that quantum mechanics itself is an approximation to a more fundamental
theory not yet known. An extremely good approximation, indeed. I eleborate
more on that below.

\textbf{3. Spectroscopy}.

This technique gives rise to the most spectacular agreement between quantum
predictions and experimental results, the precision being sometimes better
than 1 part in 10$^{9}$. This happens, e.g. in atomic spectra with visible
light (electronic transitions) or microwaves (hyperfine transitions). The
experiments of atomic spectroscopy are frequently interpreted as
measurements of the energy levels of atoms. However, that interpretation is
not necessary (although some people arg\"{u}e that it is suggested by the
formalism). We may simply assume that we are dealing with the evolution,
governed by the quantum equations, of a beam of incoming radiation
interacting with a material system (atom, nucleus, etc.). After all
spectroscopy is a particular example of scattering experiment where a light
beam is substituted for the incoming beam of particles. Both the incoming
light and the outgoing light may be usually treated as classical.

\textbf{4. Interference. }

These experiments are currently taken as the most dramatic examples of
non-classical (quantum) behaviour. Nevertheless we need rather weak
postulates for their interpretation. Again, it is enough to know the initial
state, the evolution (including the interaction of microobjects with macro
or mesoscopic devices like a grating or a detector) and the interpretation
of the final results as in collision experiments (e.g. interpretation of
what wee see in a photographic plate as blackening of grains by the action
of the incoming particles). We do not need to speak about whether a \textit{%
particle} goes through one or both slits.

In all these examples we see that the standard postulates about the
existence of discrete energy states, and their correspondence with vectors
in Hilbert space, or about the association of observables with operators,
are not really necessary. This leads us to conjecture that the
(mathematical) \emph{formalism} of quantum mechanics, the standard \emph{%
postulates of macroscopic physics} for the connection formalism-experiments,
plus some \emph{particular hypotheses} (like the assumption that an atom
consists of a nucleus plus electrons) are sufficient for the interpretation
of all experiments.

Of course, for the applications it is more economical to use some
``practical recipes'', like Feynman's rule of summing probability amplitudes
of indistinguishable paths in experiments of interference, but summing
probabilities if the paths are distinguishable. The problem is that
conceptual difficulties arise when the practical rules are taken as
postulates. For example, the wave-particle duality appears as highly
counterintuitive in experiments showing, alternatively, recombination and
anticorrelation. In a typical experiment \cite{grangier} one sends
``individual photons'' to a beam-splitter and verifies anticorrelated
detection, that is either the detector in the transmitted beam or the
detector in the reflected beam fires, but not both. This apparently proves
the corpuscular nature of the photon. However, recombination of the two
beams gives rise to interference, which apparently shows that the photon has
gone by both paths at the same time. The experiment appears as mind boggling
because it cannot be explained either assuming that something travels along
both paths or assuming that there is propagation only along one path.
However, there are alternative interpretations where something \emph{real }%
(an electromagnetic field actually) propagates by the two paths, which
explains interference, but some mechanism prevents the firing of both
detectors at the same time, which explains anticorrelated detection \cite
{ms88}.

Another crucial point of our proposed approach is that it is not necessary
to \emph{postulate} the existence of discrete energy states of quantum
systems. Such states may be just mathematical intermediates in the
calculation of the evolution. This is the case, for instance, with Fock
states of the radiation field (e.g. single photon states). The states appear
as mathematical constructs, not necessarily representing anything real, and
are similar to the Fourier components in the standard solution of linear
partial differential equations. A typical example of these is the diffusion
equation, where the Fourier components of the series solution may not be
positive definite and this does not imply that probabilities are negative,
because only the sum of the series represents a probability. (There is a
difference with the Schr\"{o}dinger equation, however, in the fact that no
theorem of positivity exists here, similar to the theorem stating that the
fundamental solution of a diffusion equation is always positive.)

The moral of our analysis is that the standard postulates of quantum
mechanics (in particular the \emph{universal }correspondence between vectors
and states) \emph{constrains }the possible interpretations of quantum
mechanics. My conjecture is that weaker postulates may allow for alternative
interpretations free from conceptual difficulties.

\section{\protect\smallskip Proposed postulates}

Firstly I admit without any change the usual formalism of quantum theory
(Hilbert space, equations of Dirac, Klein-Gordon, Maxwell, etc.)

The postulates of connection with experiments are reduced to:

1) \emph{To every physical system we associate a Hilbert space in the
standard form.}

2) \emph{To every ``preparation'' we associate a density operator}.

A preparation is a set of well specified laboratory manipulations. But I
claim that it is necessary to specify the actual operations and the full
macroscopic context. It is not enough to say, for instance, ``I take a pair
of photons of such and such properties''. We should say something like ``I
take a crystal of specified kind, cut in such or such form, at which we
direct a laser with specified properties'', etc. That is we should specify
the full macroscopic context. States are defined by the preparation like
chemical species are defined by the recipe for obtaining them in the
laboratory (either extracting them from natural products or by synthesis).
Thus I propose that\emph{\ }

2') \emph{The density operator corresponding to a preparation is the one
having maximum von Neumann entropy compatible with our knowledge about the
system}.

I do not make any distinction between ''pure states'' and ''mixtures'', but
claim that we shall treat them on the same footing. I do \textit{not }assume
that there is a physically realizable state corresponding to every vector in
Hilbert space. In this sense I reject the standard form of the \emph{%
superposition principle. }However the break which I consider for that
principle is not of the ''superselection rule'' type, but deeper. My
conjecture is that, in most cases, \textit{physical states}\emph{\ will have
a positive (nonzero) von Neumann entropy}, pure states being just
mathematical constructs. But I do not propose this condition as a postulate,
I want just to remove the postulate that the opposite is true. In any case I
do not assume that ''pure states'' provide a complete information about a
single system, but about an ensemble of systems (i. e. those corresponding
to a given preparation procedure). In this sense my proposal may be
classified within the so-called \emph{statistical interpretation }of quantum
mechanics\cite{ballentine}

\textit{I do not postulate} anything about observables (e.g. that the
possible values obtained in a measurement are the eigenvalues of some
selfadjoint operator). But I do \emph{not }claim that such statements cannot
be a part of the theory, I only claim that all statements of that kind, if
true, \emph{should be derived }from the remaining postulates. Observable is
anything that can be actually observed (measured). Therefore the observables
should be defined by a specific method of measurement. And measurement is a
physical interaction which should be studied using the remaining postulates
of quantum theory.

We should look at the process of measurement as follows. We have a system
prepared in a specified form (that is, in some ''state'' represented by a
density operator) and an experimental (macroscopic) context, also
represented by a density operator. Both the system and the context evolve in
interaction until they arrive at a final state. At the end of the
measurement we observe a \emph{macroscopic} system and this observation does
not require special postulates (in addition to those of classical physics).
Any macroscopic apparatus is in contact with the environment (it is always
an open systems) and it is possible to prove \cite{omnes} that the evolution
gives, after a long enough time, a reduced density matrix (resulting from
taking the partial trace with respect to the environment) which is diagonal
in the coordinates representation for the macroscopic variables (e.g. the
center of mass of macroscopic bodies).

Finnally we need an \textit{objectification postulate}, which is required
for a realist interpretation of quantum mechanics\emph{.} That is we must
state what elements of the theory correspond to elements of reality, in
contrast with elements of the theory which are just mathematical constructs
useful for the formulation of the theory. An example of the former is the
distance between two macroscopic bodies, an example of the latter is a
vector in a Hilbert space. For a complete theory it is necessary that every
element of physical reality has a counterpart in the theory, but I do not
assume that quantum mechanics is complete\cite{EPR}. Thus I shall not
attempt to specify the elements of the theory which correspond to every
element of reality. I shall do that for some of the elements of reality,
namely positions of bodies. Still I do not want to make claims about the
reality of quantum particles, like electrons or photons. Maybe the said
particles are just useful mathematical constructs. Consequently I shall
postulate only the \textit{objective reality} for the position of
macroscopic bodies, as follows:

3) \emph{The center of mass of any macroscopic body, or any macroscopic part
of it, has a definite position in space at every time. The probability
distribution of positions is give by Born\'{}s rule, }that is it equals the
modulus squared of the wave function in the position representation.

\emph{\ }Certainly this postulate is open to criticism. Firstly the word
macroscopic has not a sharp meaning. A solution (admittedly not very good)
is to define as macroscopic any system with mass greater than, say, one
microgram$.$ Another possible criticism is that the postulate refers
directly to macroobjects. Now, postulates about macroscopic objects may be
derived from postulates about microobjects, but the inverse process is not
trivial. Therefore finding the consequences of our three postulates for the
microscopic domain (the one most properly quantal) may be difficult or
impossible. I am aware of these problems, but in my view they are less
dramatic than the difficulties associated to CI or MWI, as commented above.
Also I suppose that the difficulties of my approach are related to the fact
that quantum mechanics is an approximation of a more fundamental theory.
This is suggested by the fact that the conceptual difficulties are
alleviated when we pass from (elementary) quantum mechanics to quantum field
theory. For instance the stationary states with sharp energy which appear in
the solution of Schr\"{o}dinger equation are rather bizarre, but it is known
that the states are neither stationary nor sharp in energy when the
interaction with the quantized radiation field is taken into account. My
conjecture is that a fundamental theory free from interpretational
difficulties will be obtained only when the unification of quantum field
theory and general relativity is achieved.

Our third postulate implies that the reduced quantum density matrix of
macroscopic variables should be interpreted as a probability distribution.
That is we propose an \emph{ignorance interpretation }of the density matrix,
which cannot be derived either in the CI or in the MWI. In fact, in both
approaches a fundamental value is attributed to ``pure states'' of physical
systems. They correspond to vectors (more correctly rays or, equivalently,
idempotent density matrices) of the Hilbert space, whilst only
non-idempotent density matrices are seen as associated to lack of
information. As is well known this leads to the impossibility of
objectification as correctly stressed by P. Mittelstaedt \cite{busch}.

In my proposal all density matrices representing actual states have an
operational meaning (including those corresponding to rays if any): They are
associated to physically realizable preparations, and they take account of
the actual \emph{information }that we have about the system. In this sense
we leave open the possibility that the information is incomplete, even about
a system represented by a ``quantum pure state'', that is we leave open the
possibility of hidden variables. Furthermore, our objectification postulate
requires that the initial information about the system to be measured is
already incomplete, although I shall not elaborate further on this point in
this paper. Another consequence of the objectification postulate is to view
decoherence\cite{Zurek} as a loss of information closely similar to the
``coarse graining'' which happens in classical physics when we average over
the degrees of freedom which are out of our control.

In my approach the concept of ``state'' has an epistemological, rather than
ontological, character. States are rather similar to the probability
distributions (ensembles) used in classical statistical mechanics. They
refer to our knowledge about the system. However, that knowledge has a
fundamentally objective character because it rests upon an objectively
defined preparation procedure. The lack of ontological commitement derives
from my rejection of statements of principle. That is I consider meaningless
expressions like ``the maximum information which may be obtained \emph{in
principle}''. The actual information is what matters.

\section{Relation with Copenhagen, many worlds and hidden variables
interpretations}

\smallskip The approach here presented has some similarities with the CI and
also with the MWI but, at a difference with these, it allows (almost
requests) for hidden variables. Let us analyze in some detail these points.

With the Copenhagen interpretation I share: 1) the emphasis on the need of
speaking always about the macroscopic context, and 2) an operational
approach to state (preparation) and observable (measurement). Indeed, I
propose to remove all postulates relating directly the elements of the
theory (electrons, photons, etc.) with actual physical objects. In this
sense the postulates about the connection formalism-experiments refer always
to a (possibly microscopic\emph{)} \emph{system plus} its\emph{\ macroscopic
context }as in the CI. In some sense the proposed postulates go farther than
Heisenberg, for whom it is nonsense to speak about \emph{trajectories }of
electrons. The proposed intepretation avoids even assuming from the start
that the electron itself (or the photon, etc.) is a \emph{real }object,
although I do not assume the opposite either. In the proposed minimal
realist interpretation the microscopic entities are assumed to be
``theoretical (human) constructs'' useful for the description of nature at
the microscopic level, although they are related to some objective reality.

However there are three sharp differences between this approach and the CI:
1) I assume that \emph{the formalism} of quantum mechanics should be applied
both to micro and to macroscopic bodies, in contrast with CI (at least with
Bohr\'{}s view), 2) I do not exclude the existence of a subquantum level
(hidden variables) which in the future might be accesible to our knowledge,
and 3) related to this is the fact that the proposed interpretation is not
considered the final word, it is just a provisional one to be used until we
have a more fundamental theory.

With the MWI (or relative state interpretation) I share the assumption of
the full validity of the quantum \emph{formalism }even for the macroscopic
world. However, the objectification postulate implies that the macroscopic
variables always possess values (all macroscopic measurements may be reduced
to position measurements). But the objectification postulate applies to a
reduced density operator, obtained by taking the partial trace with respect
to the environment. Consequently it does not apply to the whole universe,
which has no environment. Also, as mentioned above, I do not postulate any
fundamental relevance for the ``quantum pure states'', which seems to be a
(maybe implicit) assumption of the MWI.

\section{\textbf{Quantum mechanics and local realism: Bell's theorem.}}

\ It seems obvious, at least to me, that the best interpretation of quantum
mechanics would be in terms of local hidden variables, if this were
possible. (More properly, it would be desirable to have a local realist
substitute for quantum mechanics, in a similar way that general reltivity is
a local substitute for Newtonian gravitation. I believe that this was
Einstein\'{}s desideratum). Consequently Bell\'{}s theorem is the biggest
problem for a satisfactory interpretation of quantum mechanics. But I claim
that local hidden variables have not yet been excluded by performed
experiments (see below.)

The proof of Bell's theorem requires to assume that there are states such
that: 1) they are physically realizable in the laboratory, and 2) they
violate a Bell inequality . As I do not admit as a postulate of quantum
mechanics the realizability of any particular state, the derivation of the
theorem would involve proving that such state may be produced. That is I
demand a detailed prosposal of an experiment where such state may be
manufactured before a rigorous claim may be made about the incompatibility
of local realism with the empirical predictions of quantum mechanics. On the
other hand, a detailed experimental proposal is proved to be truly reliable
only when the experiment is actually made. Consequently no claim of
incompatibility may be made until such experiment is performed.

In the meantime Bell\'{}s is a purely mathematical theorem (purely means
without direct implications for the real world) that shows the
incompatibility between two formalisms: 1) the Hilbert space formalism 
\textit{plus the postulate that all vectors correspond to states which may
be physically realized }(except for the superselection rules), and 2) the
Bell formalism for local hidden variables theories. This does not mean that
Bell's theorem is irrelevant. On the contrary, I think that \emph{it is one
of the most important discoveries in the physics of the 20th century}. But
its relevance consists of being a guide for possible experiments able to
test quantum theory versus local realism. As is well known (or rather, it 
\emph{should} be well known)\emph{\ no (loophole-free) experiment has been
performed able to refute local realism up to now}. It is remarkable that
this happens more than forty years after Bell\'{}s work, which shows that
the empirical disrproof of local realism is not a trivial matter. Actually
the optical photon experiments are unable to truly test the Bell
inequalities due to the detection loophole\cite{Santos}. More suitable seem
to be experiments with atomic qubits. One such experiment has already been
performed\cite{Maryland}, but it did not close the locality loophole and
presents other difficulties\cite{Santos09}.

My conjecture is that no experiment will ever refute quantum mechanics. But
I also guess that decoherence and other sources of noise (e.g. quantum
zeropoint fluctuations) might prevent the violation of local realism. That
is, I still believe that quantum mechanics and local hidden variables are
compatible, provided we define quantum mechanics with a set of postulates
far weaker than is made usually, in the line shown in this paper.

\end{document}